\documentclass[10pt]{article}
\usepackage{lmodern}
\usepackage[letterpaper,margin=1.0in]{geometry}

\usepackage{hyperref}
\usepackage{graphicx} 
\usepackage{amsmath}
\usepackage{amsthm}
\usepackage{algorithm}
\usepackage{algpseudocode}
\usepackage{pythonhighlight}
\usepackage{amsfonts}
\usepackage{IEEEtrantools}
\usepackage{lipsum,framed}
\usepackage{tikz}
\newcommand\tikznode[3][]%
{\tikz[remember picture,baseline=(#2.base)]
    \node[minimum size=0pt,inner sep=0pt,#1](#2){#3};%
}
\usepackage[toc,page]{appendix}
\newtheorem{definition}{Definition}
\newtheorem{theorem}{Theorem}

\newtheorem{proposition}{Proposition}

\newtheorem{researchproblem}{\bf Research Problem}

\algdef{SE}[SUBALG]{Indent}{EndIndent}{}{\algorithmicend\ }%
\algtext*{Indent}
\algtext*{EndIndent}

\newcommand{\shorten}[1]{}

\begin{document}
\title{An LLM Framework For Cryptography Over Chat Channels}

\author{Danilo Gligoroski\thanks{Norwegian University of Science and Technology, Trondheim, Norway, email: danilog@ntnu.no} \and Mayank Raikwar\thanks{University of Oslo, Oslo, Norway, email: mayankr@ifi.uio.no} \and Sonu Kumar Jha\thanks{Norwegian University of Science and Technology, Trondheim, Norway, email: sonu.k.jha@ntnu.no}}



\maketitle

\begin{abstract}
 Recent advancements in Large Language Models (LLMs) have transformed communication, yet their role in secure messaging remains underexplored, especially in surveillance-heavy environments. At the same time, many governments all over the world are proposing legislation to detect, backdoor, or even ban encrypted communication. That emphasizes the need for alternative ways to communicate securely and covertly over open channels. We propose a novel cryptographic embedding framework that enables covert Public Key or Symmetric Key encrypted communication over public chat channels with human-like produced texts. Some unique properties of our framework are: 1. It is LLM agnostic, i.e., it allows participants to use different local LLM models independently; 2. It is pre- or post-quantum agnostic; 3. It ensures indistinguishability from human-like chat-produced texts. Thus, it offers a viable alternative where traditional encryption is detectable and restricted. 
\end{abstract}

\newpage
\tableofcontents
\newpage

\section{Introduction}

It is an undisputed and consensually accepted fact that Artificial Intelligence (AI) and Machine Learning are the most disruptive technologies that cause civilization and societal transformation. They profoundly affect industry, economy, work relations, and manufacturing procedures \cite{electronics12051102}. 

Large Language Models (LLMs), most notably ChatGPT~\cite{openai2022chatgpt} but also free and open source models like LLaMA~\cite{touvron2023llama2}, and DeepSeek~\cite{bi2024deepseek}, recently the top-ranked Grok 3 \cite{grok3} (in \cite{chiang2024chatbotarenaopenplatform} where more than 200 models are ranked), are the key part of why AI is seen as a disruptive technology. They're particularly impactful in human language applications like customer service, education, task automation, and efficiency improvement. While other AI fields, such as computer vision, are also disruptive, LLMs are currently at the forefront due to their recent popularity and broad applications. 
LLM models possess remarkable capabilities in generating human-like text responses and addressing queries across various aspects of daily life. Leveraging the attention mechanism, they process and generate contextually relevant responses with high coherence. LLMs can also facilitate communication between two parties, each using their own LLM to generate responses, enabling AI-mediated conversations. 

With increasing interest in LLMs, significant research~\cite{christ2024undetectable,zamir2024excuse,cryptoeprint:2024/759,zhang2024remark} has focused on embedding secret messages within LLM-generated responses, particularly for secure communication between two parties. These approaches enable private communication over public channels, as LLM responses are inherently transmitted in such settings. The security of these methods is fundamentally based on the principle that an LLM-generated response containing an embedded secret message must be indistinguishable from a regular LLM-generated response without any hidden information, preventing adversarial detection.

Nevertheless, all the previous approaches are based on certain assumptions, e.g., about the minimum entropy of the public channel or the implementation of a random oracle. These assumptions are rather strict and hard to follow while implementing these approaches. Therefore, in this work, we remove these assumptions and present a construction to enable private communication over public chat channels. To do that, we construct a special function that embeds a given set of characters in specific positions in LLM-generated text. Though there is a recent work~\cite{Conde2025} addressing if ChatGPT can count letters, the special function in our work addresses a bit similar but much harder problem of indistinguishable embedding ciphertexts at exactly desired positions. This challenge becomes more critical amid rising threats to user data privacy.
   
\subsection{Motivation}
Recent political agendas and actions significantly threaten user data privacy, as evidenced by multiple recent events. The UK government has demanded that Apple implement a backdoor to access users' encrypted data~\cite{guardian2025ukapple}. Similarly, the French government considered measures to allow message transmission within the framework of investigative requests~\cite{lemonde2025chiffrement}. In a related development, Russia-backed hacking groups have devised techniques to compromise encrypted messaging services, including Signal, WhatsApp, and Telegram~\cite{computerweekly2025signal}. These developments raise serious concerns about the future of secure communication. Given the potential scenario where public communication lacks encryption, it becomes crucial to explore alternative methods for embedding hidden information within publicly available content. This work addresses this challenge and proposes a novel approach to achieving covert communication under such constraints.

While existing techniques such as anamorphic encryption~\cite{persiano2022anamorphic} provide a means of protecting privacy, their applicability remains limited under certain constraints. Notably, if a repressive regime were to monitor, detect, and ultimately ban all conventional encryption methods, there would be a critical need for alternative ways to communicate over open channels securely. Motivated by this challenge, our work proposes a framework to facilitate covert communication under such restrictive conditions.

\subsection{Contribution} In this paper, we present a novel framework for covert encrypted communication over public chat channels. The contributions of the paper are as follows.

\begin{enumerate}
    \item We construct a novel function $\textsc{EmbedderLLM}$ that algorithmically places given characters within contextually appropriate words at specific positions in the LLM-generated response.
    \item We propose a framework where we use $\textsc{EmbedderLLM}$ to conduct symmetric LLM cryptography or public-key cryptography. 
    \begin{enumerate}
        \item As a use case example, we show how a password-based authenticated encryption scheme can be implemented to use $\textsc{EmbedderLLM}$. It embeds a secret message within the produced ciphertext encoded in an LLM-generated text.
        \item As another use case example, we show how to implement an Elliptic Curve Diffie$-$Hellman key exchange with Ephemeral keys (ECDHE) within our framework.
    \end{enumerate}
\end{enumerate}

\subsection{Paper Organization}
The remainder of the paper is organized as follows: Section~\ref{sec:related-work} reviews related work. Section~\ref{sec:transformers} provides an overview of transformers. Section~\ref{sec:EmbedderLLM} introduces the notation and details the construction of $\textsc{EmbedderLLM}$. Section~\ref{sec:LLM-Crypto} explores the LLM framework to enable LLM-based cryptography, categorizing approaches into symmetric and public-key cryptography. Finally, Section~\ref{sec:conclusion} concludes the paper and outlines promising directions for future research.

\section{Related Work} \label{sec:related-work}

\textbf{Anamorphic Encryption} 
Persiano et al.~\cite{persiano2022anamorphic} invented the notion of anamorphic encryption, which allows private communication between parties even if a dictator gets the secret keys of the parties. Technically, an anamorphic encryption scheme enables two parties, sharing a double key, to embed covert messages within ciphertexts of an established PKE scheme. This protects against a dictator who can force the receiver to disclose the PKE secret keys, but remains unaware of the existence of the double key.

Banfi et al.~\cite{banfi2024anamorphic} refine the original anamorphic encryption model, identifying two key limitations. First, the original scheme restricts double key generation to once, requiring a new key-pair after a dictator takes power, which may raise suspicion. Second, the receiver cannot determine if a ciphertext contains a covert message. To address these, Banfi et al. propose a model allowing multiple double keys per public key and enabling covert channels after key deployment. They provide constructions showing schemes like ElGamal, Cramer-Shoup, and RSA-OAEP support these robust extensions, enhancing secure communication in authoritarian regimes.

Following the work of Banfi et al., another important contribution comes from Catalano et al. ~\cite{catalano2024limits}. In this work, the authors investigate the constraints of implementing anamorphic encryption using black-box techniques, focusing on the message space size. They show that any black-box approach can only support a message space that is polynomially bounded by the security parameter, limiting its scalability. Moreover, they prove that certain stronger forms of anamorphic encryption, like asymmetric anamorphic encryption~\cite{catalano2024anamorphic}, cannot be achieved through black-box constructions. However, under specific assumptions about the underlying public-key encryption scheme, the authors demonstrate that it is possible to realize anamorphic encryption with a much larger message space, providing valuable insights into the practical limitations and possibilities of black-box anamorphic encryption.

While anamorphic encryption is effective in mitigating risks under a dictatorship, several challenges remain. For instance, the frequency of encrypted communications between two parties may arouse suspicion, prompting the dictator to take measures that disrupt private communication. Another potential issue arises if the dictator outright bans encrypted communications or anamorphic schemes hindering surveillance~\cite{cryptoeprint:2025/233}, forcing citizens to communicate publicly. In such a scenario, private exchanges between individuals would be entirely thwarted, rendering secure communication using anamorphic encryption or any sort of encryption impossible.

\textbf{Steganography} Steganography enables covert communication by embedding secret messages within carrier signals such as text, images, audio, or video. In linguistic steganography (LS), natural language text serves as the carrier, producing a steganographic (stego-) text  that conceals hidden information. The primary challenge in LS is generating stego-texts that not only encode secret messages but also maintain naturalness and fluency to avoid detection.

Wang et al.~\cite{wang2024dairstegadynamicallyallocatedintervalbased} propose DAIRstega, a linguistic steganography method that dynamically allocates coding intervals based on token conditional probabilities using a roulette wheel approach. By favoring higher-probability tokens, DAIRstega enhances the naturalness of the generated text while embedding secret messages efficiently. This approach improves stego-text quality, making it harder for adversaries to distinguish from ordinary text.

Steganographic communication typically occurs in a public setting where an eavesdropper, Eve, attempts to detect hidden messages. Alice, the sender, must ensure that Bob, the receiver, can decode the message while minimizing the risk of Eve discovering its presence. A conventional analogy compares Bob to a prisoner receiving a letter from Alice, a family member outside the prison, while Eve, the prison guard, scrutinizes the letter for unusual content. Traditional linguistic steganography methods modify an existing cover text through subtle alterations, such as synonym replacements, to avoid detection. However, with advancements in generative models, particularly LLMs, coverless steganography has emerged as a promising alternative. This approach generates stego-texts that appear indistinguishable from natural language while embedding more information within shorter messages compared to conventional cover-text-based techniques.

Huang et al.~\cite{Huang2024ODStegaLN} introduce a coverless LLM-based steganography method where an arithmetic coding decoder guides text generation to embed secret messages seamlessly. Their approach optimizes a modified probability distribution for token generation while imposing a KL divergence constraint to balance secrecy and fluency. Their work demonstrates how LLMs can be leveraged to improve the security and reliability of linguistic steganography. Building upon this, Huang et al.~\cite{huang2025relatively} introduce a framework that enhances embedding efficiency by modeling the steganographic process as a Constrained Markov Decision Process (CMDP). In recent work, Bai et al.~\cite{bai2025shifting} introduced a method that pseudorandomly shifts the probability interval of an LLM's distribution to create a private distribution for token sampling.

An LLM-based steganography has certain assumptions. For example, the tokenization process for the sender and the receiver parties must match. Another assumption is about the entropy of the channel which defines the undetectability property of the steganography. However, these assumptions can be exploited, e.g., by tokenization error, creating a drawback of LLM-steganography. 

However, a recent work~\cite{wang2025sparsampefficientprovablysecure} employs a similar idea of embedding a message within a context using sparse sampling for provably secure steganography. Nevertheless, technicalities such as optimal ranges for human-like chat conversations, embedding of the most frequent characters, cryptographic approach for defining the framework, and broadness to cover both symmetric and public-key cryptography is where we differ.

\textbf{Watermarking} Watermarking, like steganography, embeds secret messages in a model's output. However, unlike steganography, watermarking ensures the message remains detectable even after modifications. With the rise of LLMs, research has increasingly focused on developing robust watermarking techniques for generated text~\cite{christ2024undetectable,zamir2024excuse,huang2024multi,cryptoeprint:2024/759,zhang2024remark,zhang2025robust}, as detailed in recent surveys~\cite{liang2024watermarking,liu2024survey}. While watermarking is used to verify ownership, authorship, or track usage across various media, steganography prioritizes confidentiality and undetectability, concealing the the hidden message's presence.

Recent advancements in watermarking techniques for LLMs have focused on embedding undetectable watermarks to ensure the integrity and authenticity of generated text. Christ et al.~\cite{christ2024undetectable} introduced a cryptographically-inspired method where watermarks can only be detected with a secret key, making it computationally infeasible for unauthorized users to distinguish between watermarked and non-watermarked outputs. This approach ensures that the quality of the text remains unaffected, and the watermark remains undetectable even under adaptive querying.

Building upon the concept of undetectable watermarks, Zamir~\cite{zamir2024excuse} proposed a method to hide arbitrary secret payloads within LLM responses. A secret key is required to extract the payload, and without it, distinguishing between original and payload-embedded responses is provably impossible. Notably, this technique preserves the generated text quality, thereby extending the applicability of undetectable watermarking in secure communications. 

To counter adaptive users, Cohen et al.~\cite{cryptoeprint:2024/759} developed a multi-user watermarking scheme that traces model-generated text to users or groups, even under adaptive prompting. Their construction builds on undetectable, adaptively robust, zero-bit watermarking, ensuring watermark detectability despite text modifications.

To further enhance the resilience of watermarks against various text alterations, recent research~\cite{zhang2024remark,zhang2025robust} has proposed methods that maintain the watermark's detectability even after the text undergoes paraphrasing or other modifications. These approaches aim to ensure that the embedded watermarks are not only imperceptible but also robust against a range of potential attacks, thereby strengthening the security and reliability of watermarking in LLMs.

These studies collectively illustrate the evolving landscape of LLM watermarking techniques, emphasizing the need to balance undetectability, robustness, and adaptability to diverse user behaviors and adversarial threats. However, key properties such as robustness and publicly accessible detection APIs can also introduce vulnerabilities, potentially exposing these systems to various attacks~\cite{pang2024attacking,pang2024no,diaa2024optimizing}.


\section{Transformers} \label{sec:transformers}
LLMs represent a specific type of neural network architecture. They use transformers and attention mechanisms, which Vaswani et al. introduced in a 2017 paper~\cite{vaswani2017attention} titled ``Attention Is All You Need."

The fundamental mathematical techniques and theories employed in all LLMs include: 1. Linear Algebra methods such as high-dimensional embeddings and matrix operations; 2. Probability Theory, particularly the softmax functions that transform numerical values (typically the results of dot products) into probabilities (probability distributions) over tokens; and 3. Calculus, notably partial derivatives used in gradient descent algorithms to identify minima. 

\shorten{
Neural networks typically consist of layers of neurons, each with assigned weights and activation functions. The fundamental concept is that the network adjusts these weights during training to minimise a loss function using optimisation algorithms such as gradient descent. However, large language models (LLMs) represent specific neural network architectures that use self-attention mechanisms. Attention mechanisms enable the model to concentrate on various parts of the input when processing each token. For instance, when translating a sentence, the model may focus more on the subject when translating the verb. Self-attention refers to the model applying attention to its own input, allowing each position to relate to all positions in the previous layer.

A Transformer is a deep learning architecture based on self-attention mechanisms that process input sequences in parallel, making it highly efficient for tasks like natural language processing (NLP), computer vision, and time-series analysis.
} 


In this work, we will use some of the transformer's parameters known as \textit{temperature} $T$, \textit{top-$k$}, and \textit{top-$p$ (nucleus) sampling}. Those parameters are used to control the diversity and quality of the generated text. Let us briefly elaborate on the role of these parameters.

Transformers output a set of raw scores, called \textit{logits}, for each token in the vocabulary. These logits, denoted \( z_i \) for token \( i \), are converted into probabilities using the \textit{softmax function}:

\[ 
p_i = \frac{e^{z_i}}{\sum_{j} e^{z_j}} 
\]

Here, \( p_i \) represents the probability of token \( i \), and the sum is calculated across all tokens in the vocabulary. The model then samples the next token from this probability distribution. Parameters such as temperature \( T \), top-\( k \), and top-\( p \) adjust these probabilities.

Temperature \( T \), is a parameter that scales the logits before applying the softmax function. The modified probability distribution becomes:
\begin{equation}\label{Eq:Temperature}
p_i = \frac{e^{z_i / T}}{\sum_{j} e^{z_j / T}}     
\end{equation}

\begin{itemize}
    \item When \( T = 1 \): This is the standard softmax, and the probabilities reflect the model's raw predictions.  
    When \( T > 1 \): Dividing the logits by a larger \( T \) diminishes the magnitude of the exponents, making the distinctions between logits less emphasized. This flattens the distribution and leads to probabilities that are more uniform. Consequently, the model becomes less confident, fostering increased diversity in token selection.
    \item When \( T < 1 \): Dividing by a smaller \( T \) amplifies the exponents, exaggerating differences between logits. This sharpens the distribution, making high-probability tokens even more likely and low-probability ones less likely. The output becomes more deterministic and focused.
\end{itemize}
Thus, temperature controls the \textit{smoothness} or \textit{confidence} of the probability distribution: High temperature (\( T > 1 \)) invokes more randomness and creativity; Low temperature (\( T < 1 \)) causes more predictable, high-confidence outputs.

Top-$k$ sampling restricts the model to sample from only the \textit{$k$ most likely tokens}, rather than the entire vocabulary. More concretely:
\begin{enumerate}
    \item Identify the \( k \) tokens 
    \begin{equation}\label{Eq:Y_top-k-tokens}
        Y_{top-k} = \{y_1, \ldots, y_k \}
    \end{equation}
     with the highest logits, say \( z_{i_1}, z_{i_2}, \dots, z_{i_k} \) (where \( i_1, i_2, \dots, i_k \) are the indices of these tokens).
    \item Compute probabilities only over $Y_{top-k}$ tokens, setting the probabilities of all other tokens to zero: 
    \begin{equation}\label{Eq:top-k-probabilities}
        p_{i_j} = \frac{e^{z_{i_j}}}{\sum_{m=1}^{k} e^{z_{i_m}}}
    \end{equation}
    for \( j = 1, 2, \dots, k \), and \( p_i = 0 \) for all other tokens.
    \item Sample the next token from this truncated distribution.
\end{enumerate}
Thus, the effect of top-$k$ sampling is that it eliminates unlikely tokens, which reduces the chance of incoherent or random outputs. Smaller values of \( k \) create more focused output, while larger values of \( k \) allow for greater diversity among the top candidates.

Top-$p$ sampling, also known as nucleus sampling, is a technique that dynamically selects a subset of tokens based on cumulative probability instead of using a fixed number \( k \). Given a threshold \( p \) (e.g., 0.9):
\begin{enumerate}
    \item Sort the tokens by decreasing probability: \( p_{i_1} \geq p_{i_2} \geq \dots \geq p_{i_n} \).
    \item Find the smallest set of tokens (the ``nucleus'') such that their cumulative probability is at least \( p \): \[ \sum_{j=1}^{k} p_{i_j} \geq p \]
    \item Sample only from these \( k \) tokens, either by keeping their original probabilities or renormalizing them over the nucleus (depending on the implementation).
\end{enumerate}
Unlike top-$k$, which fixes the number of tokens, the effect of top-$p$ is that it adapts the size of the candidate set based on the distribution. It focuses on the most probable tokens that collectively account for \( p \) of the probability mass, balancing diversity and coherence.

Temperature and sampling methods like top-$k$ or top-$p$ are often used together. The standard approach is:
\begin{enumerate}
    \item Apply temperature to the logits to adjust the distribution: \( z_i / T \).
    \item Compute probabilities using the softmax function.
    \item Apply top-$k$ or top-$p$ sampling to truncate the distribution.
\end{enumerate}

In this work, when we write 
\begin{equation}\label{Eq:Y_top-k}
    Y_{top-k} \gets \textsc{LLM}(\textit{TOPIC}, Story, T, k)
\end{equation}
It means that we ask LLM to extend the story $Story$ within the topic $\textit{TOPIC}$;  moreover, instead of picking a token, we ask the transformer to use the temperature $T$ and to identify the top $k$ most probable tokens $Y_{top-k} = \{y_1, \ldots, y_k \}$. In the next section, we will add more criteria for selecting which token from $Y_{top-k}$ to choose.

Since this research aims to establish a framework for strong cryptographic communication over human-like texts generated by LLMs, a generic advice for the parameters is: Start with $T = 0.7$. It will produce a focused story that still has an internal variety and will mimic a human who types in a thoughtful, yet casual way. Choose $k = 40$ for producing texts with lexical variety, yet still coherent and diverse. That resembles how humans choose words from their familiar word capacity. While the generic LLM folklore advice concerning the value $p$ is to be in the range $[0.9, 0.95]$, and many LLM use cases that mimic human-like texts prefer using the top-$p$ parameter over top-$k$, in this work, we will not use and change top-$p$ parameter. 


\section{Notations and a special function $\textsc{EmbedderLLM}$} \label{sec:EmbedderLLM}

Before we outline our framework for performing cryptographic operations on LLM outputs, let us first introduce the following notations and definitions.

Let us denote the sets $S_i,\, i=1,\ldots,4$ as $S_1 = \{\texttt{0}, \texttt{1}\}$, $S_2 = \{\texttt{0}, \texttt{1}, \texttt{2}, \texttt{3}\}$, $S_3 = \{\texttt{0}, \ldots, \texttt{7}\}$, $S_4 = \{\texttt{0}, \ldots, \texttt{F}\}$, where the elements in $S_4$ are the 16 hexadecimal numbers from $0$ up to $\texttt{F}$.

In a similar manner, we denote the sets $L_i, \, i=1,\ldots,4$ as\break $$L_1 = \{\texttt{' '}, \texttt{E}\},$$ 
$$L_2 = \{\texttt{' '}, \texttt{E}, \texttt{T}, \texttt{A}\},$$ 
$$L_3 = \{\texttt{' '}, \texttt{E}, \texttt{T}, \texttt{A}, \texttt{O}, \texttt{N}, \texttt{I}, \texttt{S}\},$$ 
$$L_4 = \{\texttt{' '}, \texttt{E}, \texttt{T}, \texttt{A}, \texttt{O}, \texttt{N}, \texttt{I}, \texttt{S}, \texttt{R}, \texttt{H}, \texttt{D}, \texttt{L}, \texttt{U}, \texttt{C}, \texttt{M}, \texttt{F}\}.$$ 

We note that while $L_1 \subset L_2 \subset L_3 \subset L_4$, the set $L_4$ is a sorted set of 15 most frequent letters in English texts prepended with an even more frequent character in English texts, and which is the blank (SPACE) character. The letters' frequencies are taken from \cite{wikipediaLetterFrequency} and adjusted by adding the SPACE character as done in \cite{archiveStatisticalDistributions} and here presented in Table \ref{Tab:FrequencesTop16Characters}. The case of the letters (lowercase or uppercase) was considered equivalent. Notice the bold font probabilities for the letters $\texttt{E}$, $\texttt{A}$, $\texttt{S}$ and $\texttt{F}$.

For $i=1, \ldots, 4$, we define four bijective maps $h_i : S_i \rightarrow L_i$ where the mapping is the natural mapping between the ordered sets $S_i$ and $L_i$, i.e., $\texttt{0} \mapsto \texttt{' '}$, $\texttt{1} \mapsto \texttt{E}$, $\texttt{2} \mapsto \texttt{T}$ and so on until $\texttt{F} \mapsto \texttt{F}$. While $ h_1, h_2$, and $ h_4$ are the most adequate maps for our purposes, we note that with some technical tricks (padding the ciphertexts), we can also use the map $h_3$. 
\begin{table*}[ht!]
    \centering
    \caption{Frequencies of the top 16 characters in English texts (including SPACE) \cite{archiveStatisticalDistributions}.}
    \includegraphics[width=\textwidth]{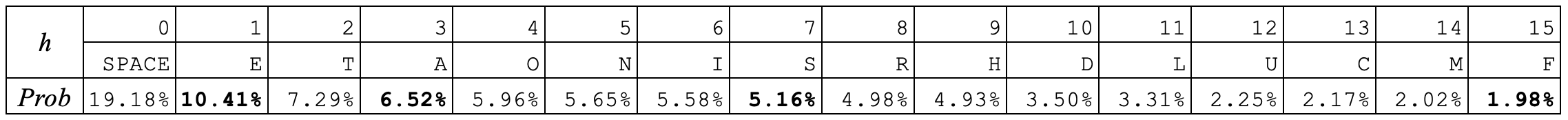}
    \label{Tab:FrequencesTop16Characters}
\end{table*}

\begin{table*}[ht!]
    \centering
    \caption{Digram frequencies of the top 16 characters in English texts (including SPACE) \cite{archiveStatisticalDistributions}.}
    \includegraphics[width=\textwidth]{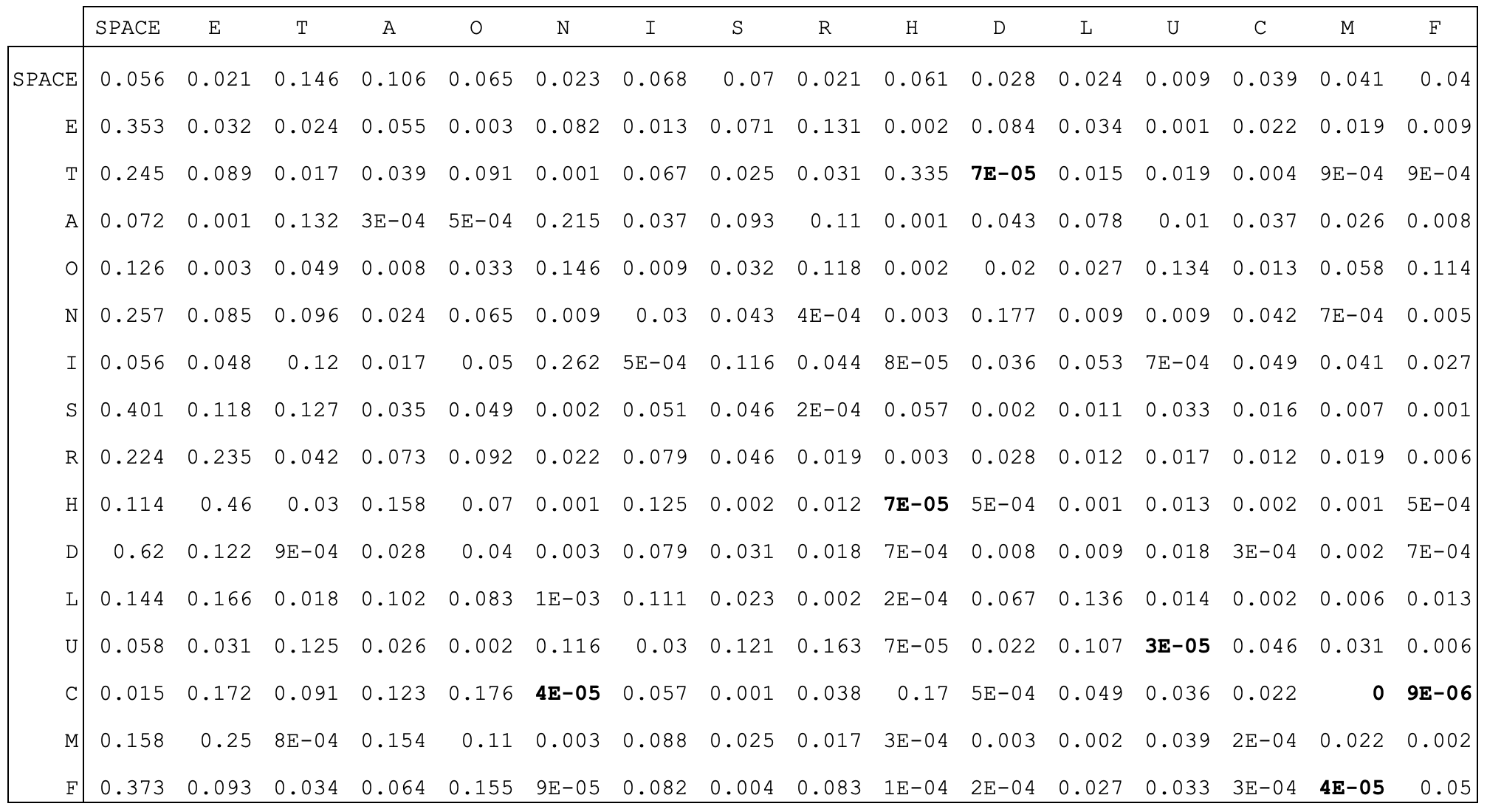}
    \label{Tab:FrequencesTop16CharactersDigramsOffset0}
\end{table*}

With a slight abuse of notation, when we write $h_i(Enc)$ for a hexadecimal string $Enc$, we refer to a character-by-character mapping of each character in $Enc$ to its corresponding image defined by $h_i$.

We will extend the generic top-$k$ sampling described with equations (\ref{Eq:top-k-probabilities}) and (\ref{Eq:Y_top-k}) with the function $\textsc{EmbedderLLM}$. This function adds additional criteria for selecting the next token from $Y_{top-k}$. The details of this function are as follows:
$$\textsc{EmbedderLLM}(\text{LLM}, \textit{TOPIC}, Story_0, T_0, k_0, \mathbf{C}, \mathbf{b})$$ 
It outputs a string $Story$. As input parameters, it receives the name of a specific LLM model to be used, a topic $\textit{TOPIC}$ discussed in the $Story$, a potentially previously generated $Story_0$ that we want to continue, a starting value $T_0$ for a temperature, an initial value $k_0$ for the top $k_0$ tokens with top $k_0$ probabilities, a sequence $\mathbf{C} = [C_0, C_1, \ldots, C_{n-1}] $ of characters from some of the sets $S_1,\ldots, S_4$, and an increasing sequence of integers $\mathbf{b} = [b_0, b_1, \ldots, b_{n-1}]$. Notice that both $\mathbf{C}$ and $\mathbf{b}$ have the same number of elements $n$.

An implicit yet crucial parameter regarding the generation of the sequence $\mathbf{b}$ is a parameter $d_o$ called the offset value. Its role is described as follows. Let us first denote by $\mathtt{PRF}()$ some pseudo-random cryptographic function that, once properly initiated, with each call, gives cryptographically strong uniformly distributed pseudo-random integer outputs in the range $[0, 2^{bit\_chunk\_size})$ for some value of $bit\_chunk\_size$. For example, if $bit\_chunk\_size = 5$, the function $\mathtt{PRF}()$ will give uniformly distributed outputs in the range $[0, 32)$. Next, when we write $b = \mathtt{PRF}()$ it means that $b$ has received a pseudo-random integer value generated by $\mathtt{PRF}()$ and $b$ is uniformly distributed in the range $b \in [0, 2^{bit\_chunk\_size})$.

Then, the sequence $\mathbf{b} = [b_0, b_1, \ldots, b_{n-1}]$ is recursively generated as follows:
\begin{equation}
    \left\{
    \begin{array}{rll}
         b_0 & \gets d_o + \mathtt{PRF}(), \\
         b_i & \gets b_{i-1} + d_o + \mathtt{PRF}(), & \text{for \ } i \in [1,n-1]
    \end{array} \right.
    \label{Eq:Sequence_b}
\end{equation}

The goal in defining $\textsc{EmbedderLLM}$ is to produce a grammatically correct and sound string $Story$ such that its length is constrained $$b_{n-1}+1 \leq \text{len}(Story) \leq b_{n-1} + d_o - 1,$$ and where the set $\mathbf{C}$ is embedded in $Story$ exactly on positions given in $\mathbf{b}$, i.e.,  $$\text{Uppercase}(Story[b_i]) = C_i,\, \text{for } i \in [0, n),$$ and the words containing characters $C_i$ are from language dictionary on which LLM was trained. In our experiments, we used LLMs trained in the English language.

Before we describe algorithmically the function $\textsc{EmbedderLLM}$, let us introduce the following additional notations. 

We overload the definition of indexing the characters of a string as follows:
\begin{equation}
    \mathtt{Char}(string, b) = 
    \left\{
    \begin{array}{@{}rll}
         string[b], & \text{if } b < len(string) \\
     \mathbf{None}, & \text{if } b \ge len(string) 
    \end{array} \right.
    \label{Eq:Overloaded_character_indexing}
\end{equation}
where $len(string)$ returns the length of $string$ and $string[b]$ is the usual 0-based Python string indexing. The reason for this overloading is that the usual Python string indexing will return $\mathtt{IndexError:\ string\ index\ out}$\\ $\mathtt{of\ range}$ if the value of $b$ is greater or equal to the length of the string.

When we write $$string \gets \varepsilon,$$ then means $string$ becomes an empty string. 

The expression $$string \gets string||token,$$ means that the new value of $string$ is the old $string$ concatenated with another string called $token$,  while the expression $$string \gets string[:position],$$ means a classical Python string slicing such that the new value of $string$ is the old $string$ stripped off by the characters from the index $position$.

If we write $$token \stackrel{\$}{\leftarrow} Y,$$ it means that $token$ gets a value from the set of tokens $Y$ and $token$ is chosen uniformly at random from $Y$. Finally the expression $$A_{\mathtt{shuffled}} \gets \mathsf{RandomShuffle}(A),$$ means the elements in $A_{\mathtt{shuffled}}$ are randomly reordered elements from $A$.

\begin{algorithm}
\caption{$\textsc{EmbedderLLM}(\text{LLM}, \textit{TOPIC}, Story_0, T_0, k_0, \mathbf{C}, \mathbf{b}, l, sec)$}
\begin{flushleft}
\textbf{Input:} LLM model, $\textit{TOPIC}$ discussed in $Story$, initial content of story $Sorty_0$, initial values for temperature $T_0$ and $k_0$ for selection of top $k$ tokens, sequence of characters $\mathbf{C} = [C_0, \ldots, C_{n-1}] $ and the sequence of integers $\mathbf{b} = [b_0,\ldots, b_{n-1}]$. The value of $l$ can be one from $[1, 2, 3, 4]$ and determines which set $L_l$ we use, and $sec \in [32, 48, 64, 96, 128]$ where $2^{-sec}$ is the probability a used token to be produced with parameters outside the optimal ranges.\\
\textbf{Output:} A text string $Story$
\end{flushleft}
 \rule{\linewidth}{0.4pt}
\begin{algorithmic}[1]
\small
\State $i \gets 0$, $n \gets len(\mathbf{C})$, $Story \gets Story_0$, $prev\_pos \gets len(Story)$
\State $Close \gets \mathtt{False}$, $T \gets T_0$, $k \gets k_0$, $Slow\_Down \gets 0$
\State $top\_f \gets Table[sec, l]$
\State $t_{\mathtt{slow\_down}} \gets \frac{0.2}{21 \times top\_f}$
\While{$i < n$}
    \State $Y_{top-k} \gets \textsc{LLM}(\textit{TOPIC}, Story, T, k)$
    \State $Y = \{y \in Y_{top-k}\ | \  \mathtt{Char}(Story || y, b_i) = C_i  \}$
    \If{$Y \neq \emptyset$} \Comment{Successful embedding}
        
        \State $next\_token \stackrel{\$}{\leftarrow} Y$
        \State $Story \gets Story || next\_token$
        \State $prev\_pos \gets len(Story)$
        \State $T \gets T_0$, $k \gets k_0$ \Comment{Reset values}
        \State $Slow\_Down \gets 0$ 
        \State $Close \gets \mathtt{False}$
        \State $i \gets i + 1$ \Comment{Go for the next embedding}

    \Else \Comment{Check if we can add an in-between token}
        \State $Y_{\mathtt{shuffled}} \gets \mathsf{RandomShuffle}(Y_{top-k})$
        \State $Unsuccessful \gets \mathtt{True}$
        \For{$next\_token \in Y_{\mathtt{shuffled}}$}
            \If{$len(Story || next\_token) < b_i - 6$}
                \State $Story \gets Story || next\_token$
                \State $Unsuccessful \gets \mathtt{False}$
                \State \texttt{Break}
            \ElsIf{$len(Story || next\_token) < b_i$}
                \If{$(\textbf{Not\ } Close)$}
                    \State $Close \gets \mathtt{True}$
                    \State $Story \gets Story || next\_token$
                    \State $Unsuccessful \gets \mathtt{False}$
                    \State \texttt{Break}
                \Else
                    \State $Slow\_Down \gets Slow\_Down + 1$
                    \If{$Slow\_Down < top\_f$}
                        \State $Unsuccessful \gets \mathtt{False}$
                        \State $T \gets T + t_{\mathtt{slow\_down}}$ \Comment{Try again. Just increase $T$}
                        \State \texttt{Break}
                    \Else
                        \State $Slow\_Down \gets 0$
                    \EndIf
                \EndIf
            \EndIf
        \EndFor
        \If{$Unsuccessful$} 
            \State $Story \gets Story[:prev\_pos]$
            \State $T \gets T + t_{\mathtt{slow\_down}}$ \Comment{Try again with increased $T$}
            \State $k \gets k + 1$ \Comment{and increased $k$\ \ \ \ \ \ \ \ \ \ \ \ \ \ \ }
            \State $Slow\_Down \gets 0$, $Close \gets \mathtt{False}$
        \EndIf
    \EndIf
\EndWhile
\State Return $Story$
\end{algorithmic}\label{Alg:EmbedderLLM}
\end{algorithm}

The crucial algorithm in our framework for LLM Cryptography is Algorithm~\ref{Alg:EmbedderLLM}. Let us discuss its steps. After the initialization steps 1 and 2, it reads a value from Table~\ref{Tab:Table} for the maximum number of repetitive attempts to find an appropriate token by increasing just the temperature before it rises the $k$ parameter. In Step 4 it calculates the temperature increase amount. Then it enters the main `while' loop for updating a human-like $Story$ about the topic \textit{TOPIC}, that contains embedded characters $\mathbf{C} = [C_0, C_1, \ldots, C_{n-1}] $ on positions $\mathbf{b} = [b_0, b_1, \ldots, b_{n-1}]$. The `while' loop is executed as long as the counter $i$ is less than $n$. In Step 6 it calls the LLM model to return a list of $k$ tokens that are candidates for continuing the $Story$. Then, in Step 7, it checks whether there are tokens $\{y\}$ that, if appended to the $Story$, would contain the character $C_i$ exactly at position $b_i$. The set of such tokens is named $Y$. In Step 8 it checks if $Y$ is a non-empty set. If yes, then in Steps 9 -- 15, it chooses uniformly at random an element of $Y$, it updates $Story$ by appending the $next\_token$, it remembers the position in $Story$ where the last successful embedding happened, it resets the values of temperature $T$, $k$, $Slow\_Down$, and $Close$ to their initial starting values (since it might happen later in the algorithm that we updated these parameters) and increases the counter $i$. If the set $Y$ is empty, the algorithm tries to append an in-between token from the initial set $Y_{top-k}$ that is randomly shuffled in Step 17. It tries all tokens from the shuffled list (Steps 19 -- 41) and checks for the first case where $len(Story || next\_token) < b_i - 6$. If that happens, it updates $Story$ with a new token (Step 21). The reason why we put the limiting value $b_i - 6$ and not $b_i$ is that we want to detect the event when the length of $Story$ will approach the crucial position $b_i$. Once entering the proximity of $b_i$ we want to try up to $top\_f$ attempts to find a non-empty $Y$ by just increasing the temperature $T$ in Step 34, before go to the part of the algorithm in Step 42. The algorithm would reach Step 42 if there was no appropriate token in all attempted $Y_{top-k}$. In that case, the algorithm shortens the story to the length where the last successful embedding happened (Step 43), relaxes both parameters $T$ and $k$ (Steps 44 and 45), and resets the variables $Slow\_Down$ and $Close$. In such a case, the `while' loop continues with the relaxed parameters and tries again to embed $C_i$ at the position $b_i$.

First, let us justify using a non-zero value $d_o$ in (\ref{Eq:Sequence_b}). Let the sequences $\mathbf{C} = [C_0, C_1, \ldots, C_{n-1}] $ and $\mathbf{b} = [b_0, b_1, \ldots, b_{n-1}]$ be the sequences produced as described above, where $\mathtt{PRF}()$ in the recurrent equations (\ref{Eq:Sequence_b}) gives uniformly distributed integers in the range $b \in [0, 2^{bit\_chunk\_size})$. 

Table \ref{Tab:FrequencesTop16CharactersDigramsOffset0} gives the probability of diagrams in English texts. Apparently, there are digrams with zero or very low probability. We can look at digrams as a pair of characters with an offset $d_o = 0$. Then, in a sufficiently long sequence of characters, we might have the situation where $(C_{i}, C_{i+1})$ is a diagram with zero or very low probability. From Table \ref{Tab:FrequencesTop16CharactersDigramsOffset0} such digrams could be $\texttt{CM}$, $\texttt{CF}$, $\texttt{CN}$, $\texttt{TD}$, $\texttt{HH}$ 
and so on (presented in bold font in the table). It means that regardless of how many calls to $\textsc{LLM}(\textit{TOPIC}, Story, T, k)$ we make in Step 5 of Algorithm \ref{Alg:EmbedderLLM}, there will be no tokens that give non-empty set $Y$ in Step 6. From an empirical perspective, in our experiments, we noticed that offset digrams where $d_o < 6$ still give potentially small probabilities. Thus, we set the default value $d_o = 32$.

\subsection{Modeling the Distinguishability of Texts Outputs From Algorithm \ref{Alg:EmbedderLLM} From Human-Like Produced Texts}

We are going to model the distinguishability of texts generated by Algorithm \ref{Alg:EmbedderLLM} from human-like produced texts within the following plausible constraints (\textit{assumptions}):
\begin{enumerate}
    \item The average token size \cite{openai2023tokens} (in number of characters) used in modern LLMs is 4.
    \item Optimal parameters for the imitation of human-produced chat texts are $T \in [0.7, 0.9]$ and $k \in [40, 60]$.
    \item \label{assumption3} We assume that the adversary possesses the exact local LLM used in the call of Algorithm \ref{Alg:EmbedderLLM}. By this, we assume that the adversary knows even the local fine-tunings of a publicly available LLM model. This makes the adversary a very powerful entity, which in practice might not be the case, but from a security perspective, it is a preferred modeling style.
    \item The adversary does not know the values of the sequences $\mathbf{C}$ and $\mathbf{b}$.
    \item Knowing the exact LLM, the adversary successfully distinguishes the captured $Story$ if it distinguishes a token in $Story$ that was produced with parameters $T$ and $k$ outside of the ranges $[0.7, 0.9]$ and $[40, 60]$ (we call these ranges "optimal ranges").
\end{enumerate}

\begin{table}[ht!]
    \centering
    \caption{A table of top-$f$ parameters for achieving probabilities less than $2^{-sec}$ for a used token to be produced with parameters outside the optimal ranges.}
    \includegraphics[width=9cm]{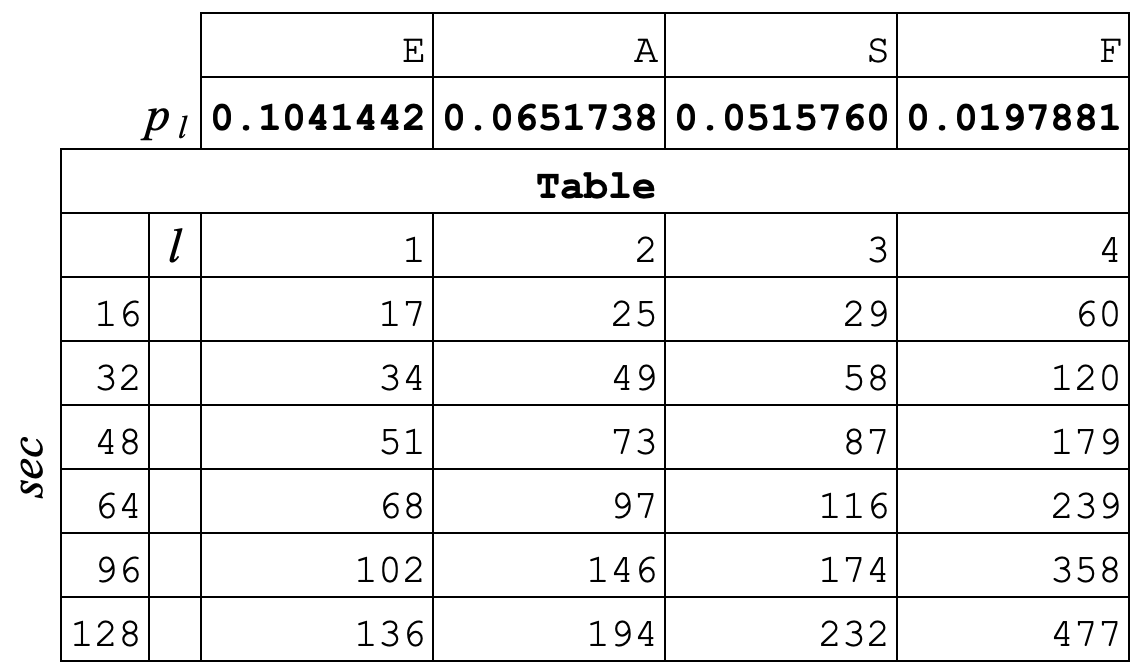}
    \label{Tab:Table}
\end{table}

\begin{theorem}\label{Thr:DistinguishinProbabilities}
The probability a $Story$ generated by  $\textsc{EmbedderLLM}(\text{LLM}, \textit{TOPIC}, \varepsilon, T_0, k_0, \mathbf{C}, \mathbf{b}, l, sec)$ to have a token produced with parameters $T\notin [0.7, 0.9]$ and $k \notin [40, 60]$ denoted as $P_{dist}$ is less than $2^{-sec}$.
\end{theorem}
\begin{proof}
    Let us compute the probability of Algorithm \ref{Alg:EmbedderLLM} to reach and execute steps 44 -- 46.
    First, let us use the fact that the average token size \cite{openai2023tokens} in modern LLMs is four characters. Plausible modeling of $k$ tokens whose average size is four is a multinomial distribution with a central value of length 4.
    More concretely, we can model that $k$ tokens belong to seven categories (seven bins) determined by token length. We consider the probability of having a token of length eight or higher to be very small. In that case, the probability that a token has a length $j \in [1,\ldots,7]$ can be modeled with 

    \begin{equation}
        P(j) = 2^{-6} \binom{6}{j-1}, \quad j \in [1,\ldots,7].    
    \end{equation}
     
    Then, in Step 7, we compute the set\\ $Y = \{y \in Y_{top-k}\ | \  \mathtt{Char}(Story || y, b_i) = C_i  \}$. The lower bound of the probability the set $Y$ is non-empty depends on the number $k$ of elements in the set $Y_{top-k}$ and the probability $p_l$ of character with the lowest frequency in the set $L_l, l\in [1,\ldots,4]$ (given in Table \ref{Tab:Table}). The lower bound can be computed as: 
    \begin{equation}
        \Pr(Y \neq \emptyset) = P(k, p_l) = \frac{1}{7} \sum _{j=1}^7 \left(\frac{p_l}{j}\right)^{k P(j)}.
        \label{Eq:YNonEmpty}
    \end{equation}

    Now we can compute the upper bound of the probability of a token to be produced with parameters $T$ and $k$ that are outside the ranges $[0.7, 0.9]$ and $[40, 60]$ as a consecutive product of the probabilities of the opposite event, i.e., 
    \begin{equation}
        \Pr(Y = \emptyset) = 1 - \Pr(Y \neq \emptyset).
    \end{equation} 
    However, here, with a slight increase in the temperature (between the events when we also increase the value of $k$) in Step 34, we want to increase the number of attempts to find a non-empty set $Y$. The number of such attempts is also given in Table \ref{Tab:Table}, and plays a role in Step 4, where we calculate the appropriate value for temperature increase $t_{\mathtt{slow\_down}}$.
    Then, the upper bound of the probability $P_{dist}$ can be  calculated with the following formula:
    \begin{equation}
        P_{dist} \leq \prod _{k=40}^{60} \left(1-P(k, p_l)\right)^{top\_f}.
    \end{equation}
    For concrete values of $l$, i.e., $p_l$ one can compute that indeed for the values $top\_f$ given in Table \ref{Tab:Table}, the probabilities $P_{dist}$ are upper bounded by $2^{-sec}$.
\end{proof}

\subsection{Discussion on Algorithm \ref{Alg:EmbedderLLM}}
\subsubsection{Execution (in)efficiency}
Algorithm \ref{Alg:EmbedderLLM} is not an efficient algorithm. It repeatedly calls the LLM model with slightly changed parameters in an effort to build a story with the required properties. A smarter strategy might involve the use of dictionaries, thesaurus, and grammar tools in order to replace certain words with appropriate synonyms or similar phrases that would not change the narrative and meaning but would tweak the length of the story such that the required character $C_i$ is placed exactly on position $b_i$. However, in that case, the probabilities modeling that we have conducted in Theorem \ref{Thr:DistinguishinProbabilities} might not hold.

\subsubsection{Redundancy (in)efficiency}
From the redundancy perspective, apparently, for sending a short encrypted text, Algorithm \ref{Alg:EmbedderLLM}, embeds it in a longer $Story$. Choosing bigger values for $d_o$ linearly increases the length of produced $Story$. Our default recommended value is $d_o=32$. It allows every token that contains a mapped ciphertext character $C_i$ to be at a safe distance from each other such that finding appropriate tokens is a procedure independent of the embedding token found previously. Another crucial parameter that affects the redundancy, i.e., the size of $Story$, is the parameter $l$, i.e., the mapping set $L_1$, $L_2$, $L_3$, or $L_4$. With smaller $l$, we get bigger values of $n$; thus, the length of $Story$ increases.

\subsubsection{Practical vs cryptographically strong probabilities} 
Table~\ref{Tab:Table} displays values of $top\_f$ for different combinations of $sec$ and $l$. However, from a cryptographic point of view, only the lowest row with $sec = 128$ offers parameters that ensure the probabilities are lower than $2^{-128}$ for producing a token with temperature $T$ and $k$ outside the optimal ranges. Why do we give parameters for lower probabilities? The answer can be summarized as: in practice, users might choose relaxed parameters and still be comfortable that all produced tokens, with overwhelming probabilities, will be within the optimal ranges. For example, in Steganography for distinguishing between classes, so-called Linear Distinguishing Analysis (LDA) is used, and typically, the values of $\Delta\text{LDA}$ (considered as good indistinguishing properties) are in the range $[0.004, 0.007]$ (see for example~\cite{wang2024dairstegadynamicallyallocatedintervalbased}). In Table~\ref{Tab:Table}, the parameters for $sec=16$ already surpass that range because $2^{-16} \approx 0.000015$.

\subsubsection{(In)Plausability of assumption ~\ref{assumption3}}
In assumption ~\ref{assumption3}, we assume that the adversary possesses the exact local LLM used in the call of Algorithm~\ref{Alg:EmbedderLLM}. In practice, this makes the adversary a very strong adversary. For example, currently, on ``Hugging Face" there are 1,514,210 models \cite{huggingfaceModelsHugging}. Moreover, users can locally produce their own variants of the models, and those models can be kept private. 

\subsubsection{Using Algorithm \ref{Alg:EmbedderLLM} with more than one LLM model}
Closely related to the previous point, users can modify the algorithm so that in Step 6, it calls different LLM models. Users can even call several different models in parallel, check which one gives non-empty $Y$, and continue to the next $C_i$ and $b_i$.


\section{LLM Cryptography} \label{sec:LLM-Crypto}

\subsection{Symmetric Key LLM Cryptography (Password-Based AEAD)}

\textbf{Key Idea:}
Alice and Bob want to have a secure communication. They share an initial secret password. They generate keys $dk_1, dk_2$ using a Password-based Key Derivation Function (PBKDF2)~\cite{rfc8018}. To send a secret message $M=plaintext$, Alice encrypts it using Authenticated Encryption with Associated Data (AEAD) function~\cite{Black2025} with key $dk_1$, generating ciphertext $Enc$. She then maps $Enc$ to an encoded form $C$ using a function $h$, ensuring the characters belong to a set of frequent English letters. Alice embeds $C$ in a generated \textit{Story} on a \textit{Topic} using $\textsc{EmbedderLLM}$, which algorithmically places each character of $C$ within contextually appropriate words at specific positions, determined using $dk_2$ and public parameters. The crucial challenge is ensuring these insertions maintain the natural distribution of human-like but LLM-generated text. Alice then transmits the story over a public channel to Bob. Upon receiving it, Bob extracts $C$ using $dk_2$, reverses the mapping via $h^{-1}$, and decrypts the recovered ciphertext using AEAD with $dk_1$ to retrieve $M$.

\textbf{Construction:}

Let us denote a generic Authenticated Encryption with Associated Data (AEAD)~\cite{Black2025} function as  $$ciphertext, tag \gets {\textsc{AEAD}}_{enc}(dk_1, nonce, AD, plaintext),$$ and its corresponding inverse function of decryption and verification as 
$$plaintext \gets {\textsc{AEAD}}_{dec}(dk_1, nonce, AD, ciphertext, tag).$$

The ${\textsc{AEAD}}_{enc}()$ function receives as parameters a secret key $dk_1$, publicly known values of $nonce$ (which should be some non-repeatable value) and associated data $AD$, and a $plaintext$ to be encrypted. It returns the encrypted output $ciphertext$ and a verification (checksum) value $tag$. The ${\textsc{AEAD}}_{dec}()$ might return the $plaintext$ if the verification tag $tag$ passes the test, or it might return the value $Fail$ otherwise. 

Let us further denote a generic Password Based Key Derivation Function (PBKDF) (version 2)~\cite{rfc8018} as $$DK = \textsc{PBKDF2}(password, Salt, count, dkLen).$$ The $\textsc{PBKDF2}()$ function receives as parameters a value of $password$, a string known as $Salt$, the number $count$ of applications of some cryptographic hash function (not explicitly mentioned here, but in practical implementations should be instantiated with functions such as SHA2 or SHA3), and the size of the output in bytes as a number $dkLen$. For example, if we put $dkLen = 64$, the output $DK$ will be long $64 \times 8 = 512$ bits.

The assumption is that Alice and Bob share a secret $password$ along with other information that need not be secret. That non-secret information is:
\begin{enumerate}
    \item Which character mapping function $h_1, \ldots, h_4$ they will use; In the descriptions below we assume they use $h_4$.
    \item Values of $count$ and $Salt$ for $\textsc{PBKDF2}$;
    \item Values of $nonce$ and $AD$ for $\textsc{AEAD}()$ functions;
    \item An offset value $d_o$. As a default value, we set $d_o = 32$.
    \item The authentication tag size in $\textsc{AEAD}()$ functions is fixed to 128 bits, i.e., to 32 hexadecimal values.
    \item The use of $\textsc{SHAKE128}$ Extendable-Output Function
    \item Parsing the output of $\textsc{\textsc{SHAKE128}}$ into chunks of $chunk\_size$ bits. As a default value, we set $chunk\_size = 5$.
\end{enumerate}

The second derived key $dk_2$ is used as a key material to initialize Extendable-Output Function (XOF) \textsc{SHAKE128} \cite{Dworkin2015SHA3SP} by calling $Init(\textsc{SHAKE128}(dk_2))$. If $\textsc{SHAKE128}$ is called again as $$ \textsc{SHAKE128}(chunk\_size), $$ its output can be seen as an $chunk\_size$ bits output from the XOF \textsc{SHAKE128}. 

\begin{algorithm}[t]
\caption{LLM Authenticated Encryption}
\begin{flushleft}
\textbf{Input:} A shared secret $password$\\
\textbf{Output:} A text string $Story$
\end{flushleft}
 \rule{\linewidth}{0.4pt}
\begin{algorithmic}[1]
\State $dk_1, dk_2 \gets \textsc{PBKDF2}(password, Salt, count, 64)$ \Comment{\text{assert} $\text{len}(dk_1) = \text{len}(dk_2) = 256$ bits}
\State Alice generates a message \textit{plaintext} to be encrypted.
\State {\small $ciphertext, tag \gets {\textsc{AEAD}}_{enc}(dk_1, nonce, AD, plaintext)$}
\State Treat $ciphertext$ and $tag$ as hexadecimal strings.
\State $Enc \gets tag \parallel ciphertext$ 
\State $\mathbf{C} \gets h_4(Enc)$, $n \gets \text{len}(\mathbf{C})$
\State $Init(\textsc{SHAKE128}(dk_2))$
\State $b_0 = d_o + \textsc{SHAKE128}(chunk\_size)$
\State $\mathbf{b} \gets [b_i \mid i \in [0, n)]$\par \!\!\!\!\!\!\!\!\!\!\!\!\!\! where $b_i \gets b_{i-1} + d_o + \mathsf{SHAKE128}(chunk\_size)$
\State Alice chooses a topic \textit{TOPIC}.
\State $Story \gets \textsc{EmbedderLLM}(\text{LLM}, \textit{TOPIC}, \varepsilon, T_0, k_0, \mathbf{C}, \mathbf{b}, l, sec)$ 
\State Alice sends $Story$ to Bob.
\end{algorithmic}
\end{algorithm}



For authenticated decryption and verification, Bob does not require the same LLM model as Alice. In fact, he does not need any LLM model at all.

\begin{algorithm}[ht]
\caption{LLM Authenticated Decryption and Verification}
\begin{flushleft}
\textbf{Input:} A shared secret $password$ and a text string $Story$\\
\textbf{Output:} $plaintext$ or $Fail$ 
\end{flushleft}
 \rule{\linewidth}{0.4pt}
\begin{algorithmic}[1]
\State $dk_1, dk_2 \gets \textsc{PBKDF2}(password, Salt, count, 64)$
\State $Enc \gets \text{empty string}$
\State $Init(\textsc{SHAKE128}(dk_2))$
\State $pos \gets d_{o} + \textsc{SHAKE128}(chunk\_size)$
\While{$pos < \text{len}(Story)$}
    \State $Enc \gets Enc + Story[pos]$
    \State $pos \gets pos + d_{o} + \textsc{SHAKE128}(chunk\_size)$
\EndWhile
\State $tag \gets Enc_0,\ldots,Enc_{31}$
\State $tag \gets h_4^{-1}(tag)$
\State $ciphertext \gets Enc_{32},\ldots,Enc_{last}$
\State $ciphertext \gets h_4^{-1}(ciphertext)$
\State $plaintext \gets {AEAD}_{dec}(dk_1, nonce, AD, ciphertext, tag)$
\end{algorithmic}
\end{algorithm}

Notice the while loop steps 5 -- 8 in the decryption algorithm. While Alice does not send the explicit length of $tag$ and $ciphertext$, Bob, from the knowledge of $password$ and the length of $Story$, can exactly determine $tag$ and $ciphertext$.

By embedding AEAD-encrypted messages in LLM-generated text, it remains indistinguishable from normal output, even against advanced ``LLM'' or ``ML classifier'' adversaries trained on data consisting of normal/covert stories. The message $M$ is first encrypted with key $dk_{1}$, then $\textsc{EmbedderLLM}$ places the ciphertext in a fluent story guided by $dk_{2}$. Randomized token selection and natural phrasing minimize statistical traces, while PBKDF2-derived keys resist brute-force attempts. Without $dk_{1}$ and $dk_{2}$, attackers cannot detect or recover hidden data, ensuring resilience against steganalysis.

\subsection{Public Key LLM Cryptography}
\begin{algorithm}[!ht]
	\caption{ECDHE-LLM}
	\begin{flushleft}
		\textbf{Input:} Shared parameters for $\texttt{Curve25519}$. Potentially a shared secret $rootkey$. Shared parameters for operations related to LLMs such as $l$, (here $l=4$), $T_0$, $k_0$.\\
		\textbf{Output:} A shared session key $SharedKey$
	\end{flushleft}
	\rule{\linewidth}{0.4pt}
	\begin{algorithmic}[1]
            \scriptsize
		\Statex Alice and Bob compute
		\Indent
		\If{$rootkey$}
		\State $dk_1, dk_2 \gets \textsc{PBKDF2}(rootkey, Salt, count, 64)$ 
		\Else
		\State $dk_1, dk_2 \gets \textsc{PBKDF2}(\varepsilon, Salt, count, 64)$ 
		\EndIf
		\Statex \ \ \ \ \ \ \textbf{assert} $\text{len}(dk_1) = \text{len}(dk_2) = 256$ bits
		\State $Init(\textsc{SHAKE128}(dk_1))$
		\State $b_0 = d_o + \textsc{SHAKE128}(chunk\_size)$
		\State $\mathbf{b_{Alice}} \gets [b_i \mid i \in [0, 64)$\par \!\!\!\! where $b_i \gets b_{i-1} + d_o + \mathsf{SHAKE128}(chunk\_size)$
		
		\State $Init(\textsc{SHAKE128}(dk_2))$
		\State $b_0 = d_o + \textsc{SHAKE128}(chunk\_size)$
		\State $\mathbf{b_{Bob}} \gets [b_i \mid i \in [0, 64)$\par \!\!\!\! where $b_i \gets b_{i-1} + d_o + \mathsf{SHAKE128}(chunk\_size)$
		\State $Chat \gets \varepsilon$
		\EndIndent
		\Statex
		\Statex Alice computes 
		\Indent
		\State A random integer in the range $a \stackrel{\$}{\leftarrow} [1, 2^{255} - 19)$
		\State $comp_{Alice} \gets \texttt{rand\_low\_bound\_fixed\_len\_comp}(64, 3, 10)$
		\Statex \ \ \ \ \ \ \textbf{assert} $comp_{Alice} = [aa_0, aa_1, \ldots, aa_{9}]$ such that $\sum_{j=0}^{9}aa_j = 64$
		\State $ x_A = \texttt{Curve25519}(a, x_g)$
		\State $\mathbf{C}_A \gets h_4(x_A)$
		\State $\mathbf{CPart}_A \gets \texttt{Partition}(\mathbf{C}_A, comp_{Alice})$
		\State $\mathbf{bPart_{A}} \gets \texttt{Partition}(\mathbf{b_{Alice}}, comp_{Alice})$
		\State $Story_A \gets \varepsilon$
		\EndIndent
		\Statex Bob computes
		\Indent
		\State A random integer in the range $b \stackrel{\$}{\leftarrow} [1, 2^{255} - 19)$
		\State $comp_{Bob} \gets \texttt{rand\_low\_bound\_fixed\_len\_comp}(64, 3, 10)$
		\Statex \ \ \ \ \ \ \textbf{assert} $comp_{Bob} = [bb_0, bb_1, \ldots, bb_{9}]$ such that $\sum_{j=0}^{9}bb_j = 64$
		\State $ x_B = \texttt{Curve25519}(b, x_g)$
		\State $\mathbf{C}_B \gets h_4(x_B)$
		\State $\mathbf{CPart}_B \gets \texttt{Partition}(\mathbf{C}_B, comp_{Bob})$
		\State $\mathbf{bPart_{B}} \gets \texttt{Partition}(\mathbf{b_{Bob}}, comp_{Bob})$
		\State $Story_B \gets \varepsilon$
		\EndIndent
		\Statex
		\State $Chat \gets Chat || AliceInitialMessage$ 
		\State $Chat \gets Chat || BobInitialResponse$ 
		\For{$j \in [0,9]$}
		\Statex \ \ \ \ \ \ Alice side
		\State $OldStory_{A} \gets Story_{A}$
		\State $Story_{A} \gets \textsc{EmbedderLLM}(\text{LLM}, Chat, Story_{A}, T_0, k_0, \mathbf{C}_{A_j}, \mathbf{bA}_j, l, sec)$
		\State $AliceInput_j \gets Story_{A} \setminus OldStory_{A}$
		\State $Chat \gets Chat || AliceInput_j$
		\Statex \ \ \ \ \ \ Bob side
		\State $OldStory_{B} \gets Story_{B}$
		\State $Story_{B} \gets \textsc{EmbedderLLM}(\text{LLM}, Chat, Story_{B}, T_0, k_0, \mathbf{C}_{B_j}, \mathbf{bB}_j, l, sec)$
		\State $BobInput_j \gets Story_{B} \setminus OldStory_{B}$
		\State $Chat \gets Chat || BobInput_j$    
		\EndFor
		\Statex Alice side
		\State \ \ \ \ \ \ $Story_{B}\gets BobInput_0 || \dots || BobInput_9$
		\State \ \ \ \ \ \ $PublicKeyBob \gets \{ C_i \ | \ Story_{B}[\mathbf{b_{Bob}}[i]] = C_i, i\in [0,64)\}$
		\State \ \ \ \ \ \ $PublicKeyBob \gets h_4^{-1}(PublicKeyBob)$
		\State \ \ \ \ \ \ $SharedKeyAlice \gets \texttt{Curve25519}(a, PublicKeyBob)$
		\Statex Bob side
		\State \ \ \ \ \ \ $Story_{A}\gets AliceInput_0 || \dots || AliceInput_9$
		\State \ \ \ \ \ \ $PublicKeyAlice \gets \{ C_i \ | \ Story_{A}[\mathbf{b_{Alice}}[i]] = C_i, i\in [0,64)\}$
		\State \ \ \ \ \ \ $PublicKeyAlice \gets h_4^{-1}(PublicKeyAlice)$
		\State \ \ \ \ \ \ $SharedKeyBob \gets \texttt{Curve25519}(b, PublicKeyAlice)$
		\State $SharedKey = SharedKeyAlice = SharedKeyBob$
		\State $rootkey \gets SharedKey$
		\State Return $SharedKey$
	\end{algorithmic}\label{Alg:ECDHE-LLM}
\end{algorithm}

Before we describe how our $\textsc{EmbedderLLM}$ algorithm can be used in Public Key Cryptography, let us first prove one simple (although a negative) result concerning a distinguishability of texts $Story$ produced by LLMs and $\textsc{EmbedderLLM}$ whether they contain or do not contain embedded cryptographic ciphertext.

\begin{proposition}\label{Prop:SimpleDistinguisher}
Let $Story_1$ be generated by an LLM and $Story_2$ be generated by $$\textsc{EmbedderLLM}(\text{LLM}, \textit{TOPIC}, \varepsilon, T_0, k_0, \mathbf{C}, \mathbf{b}, l, sec).$$ If the sequence $\mathbf{b}$ is known to an adversary, then there is an efficient algorithm that distinguishes $Story_1$ and $Story_2$.
\end{proposition}
\begin{proof}
    It is just a simple character extraction. Let us construct two sets $Y_1 = \{C_i | \  Story_1[b_i] = C_i  \}$ and $Y_2 = \{C_i | \ Story_2[b_i] = C_i  \}$. Since $Story_1$ is generated by an LLM without any constraints, with an overwhelming probability 
    $$Y_1 \not \subseteq L_i, \; \forall i \in  [1, \ldots,4].$$

    On the other hand, we get that $$Y_2 \subseteq L_l,$$ for some $l \in [1, \ldots,4]$. 
\end{proof}

In a public key setup, the positions of characters $\mathbf{b}$ should be known to both key exchange parties. This implies that $\mathbf{b}$ should be publicly known information accessible also to the adversary. While this might seem like a negative result, there are a lot of use cases and cryptographic protocols that separate the phases of initial (covert exchange of information) key exchange and public key DH key exchange for generating session keys. The initial exchange of some information between communicating parties, which is unavailable to the adversary, is a plausible assumption. The plausibility comes from the fact that in modern society, there are numerous possibilities to perform that first phase of the initial covert information exchange, such as via mobile networks or personal physical meetings combined with websites that dynamically publish data or randomness beacons~\cite{raikwar2022sok}. We list two cases that use this plausible assumption: 
\begin{enumerate}
    \item The popular messenger application Signal \cite{signalSignalOutside} uses the ``Double Ratchet Algorithm''. It needs a root key that is subsequently used to renew and generate short-lived session keys. The session key generation combines Diffie$-$Hellman key exchange (DH) and a key derivation function (KDF). The root key is assumed to be in possession of both parties and is outside the definition of the Double Ratchet Algorithm. It can be in the form of so-called ``pre-keys'' or established with so-called ``triple Diffie$-$Hellman key exchange (3-DH)''. 
    \item The anamorphic encryption \cite{persiano2022anamorphic, banfi2024anamorphic} assumes that ``double key'' $dk$ is somehow sent covertly from Bob to Alice without the knowledge of the adversary. 
\end{enumerate}

In our case, once that initial information is shared between Alice and Bob, it can be used to produce sequences $\mathbf{b}$ unknown to the adversary. Yet, we formulate as a challenging open problem the following:
\begin{researchproblem}\label{OpenProb:KnownPositions}
Let $Story_1$ be generated by an LLM. Is it possible to design an algorithm $$ \textsc{EmbedderLLM}(\text{LLM}, \textit{TOPIC}, Story_0, T_0, k_0, \mathbf{C}, \mathbf{b}, l, sec)$$ that produces $Story_2$, such that when the sequence $\mathbf{b}$ is known to an adversary $\mathcal{A}$, then the advantage of the adversary $\mathcal{A}$ of distinguishing $Story_1$ and $Story_2$ is negligible?
\end{researchproblem}

Let us now describe an Elliptic Curve Diffie$-$Hellman key exchange with Ephemeral keys (ECDHE) within our framework. 
We will use one popular and standardized group of elliptic points, Curve25519. It is defined with the equation $$y^2 = x^3 + 486662 x^2 + x,$$ over the finite field $\mathbb{F}_{2^{255} - 19}$. As the generator point $G = (x_g, y_g)$, we take the standard value $G = (9, y_g)$, i.e., $x_g = 9$ where the value of $y_g$ is not necessary to know since it can be computed from the value of $x_g$. When we write $$ x_A = \texttt{Curve25519}(a, x_g),$$ we mean the $x$ coordinate of the multiplication of $G$ by an integer $a$ that is in the range $[1, 2^{255} - 19)$, $$(x_A, y_A) = a\cdot G.$$ We can interpret $x_A$ as a $256$-bit string (to be used with mapping $h_1$) or as a sequence of $128$ elements in the range $[0..3]$ (appropriate for the use with the mapping $h_2$) or as e sequence of $64$ hexadecimal values (in the range $[0..F]$) (appropriate for the use with the mapping $h_4$).


When we write $$comp \gets \texttt{rand\_low\_bound\_fixed\_len\_comp}(n, low, m),$$ it means that $comp$ is a list of $m$ integers greater or equal to $low$ that is a composition of $n$, i.e., $\texttt{sum}(comp) = n$. In the Algorithm \ref{Alg:ECDHE-LLM}, we use arbitrary parameters $low = 3$ and $m = 10$. That means the instruction $$comp_{Alice} \gets \texttt{rand\_low\_bound\_fixed\_len\_comp}(64, 3, 10),$$ randomly generates a list $$comp_{Alice} = [aa_0, aa_1, \ldots, aa_{9}]$$ such that $\sum_{j=0}^{9}aa_j = 64.$ Similarly, we produce the list $$comp_{Bob} = [bb_0, bb_1, \ldots, bb_{9}]$$ such that $\sum_{j=0}^{9}bb_j = 64.$

The instructions 
\begin{align*}
    \mathbf{CPart}_A & \gets \texttt{Partition}(\mathbf{C}_A, comp_{Alice}),\\ 
    \mathbf{bPart_{A}} & \gets \texttt{Partition}(\mathbf{b_{Alice}}, comp_{Alice}),
\end{align*}
partition $\mathbf{C}_A$ and $\mathbf{b_{Alice}}$ into sublists where $\mathbf{CPart}_A = [\mathbf{C}_{A_0}, \ldots, \mathbf{C}_{A_9}]$ and $\mathbf{bPart_{A}} = [\mathbf{bA}_0,\ldots, \mathbf{bA}_9]$ according to the composition $comp_{Alice}$ i.e., $len(\mathbf{C}_{A_j}) = len(\mathbf{bA}_j) = aa_j$. Similar applies for \begin{align*}
    \mathbf{CPart}_B & \gets \texttt{Partition}(\mathbf{C}_B, comp_{Bob}),\\ 
    \mathbf{bPart_{B}} & \gets \texttt{Partition}(\mathbf{b_{Bob}}, comp_{Bob}),
\end{align*}
where $\mathbf{CPart}_B = [\mathbf{C}_{B_0}, \mathbf{C}_{B_1}, \ldots, \mathbf{C}_{B_9}]$, $\mathbf{bPart_{B}} = [\mathbf{bB}_0,\ldots, \mathbf{bB}_9]$ and $len(\mathbf{C}_{B_j}) = len(\mathbf{bB}_j) = bb_j$.

By asserting that $Story_{A} \equiv OldStory_{A} || AliceInput_j$ with the instruction $$AliceInput_j \gets Story_{A} \setminus OldStory_{A},$$ we extract the new text $AliceInput_j$ that was appended to $OldStory_{A}$.

A brief discussion about the steps in Algorithm \ref{Alg:ECDHE-LLM} follows.

Alice and Bob share parameters for $\texttt{Curve25519}$. If they had a previously shared secret key (denoted as $rootkey$), they might use it to calculate the embedding positions. They also share parameters for LLM operations, such as $l$ (a default value is $l=4$), $T_0$, and $k_0$. The first 12 steps calculate the positions where the characters related to the ephemeral public keys of Alice and Bob will be embedded. Then, in steps 13--19, Alice computes her public key $x_A$, which is mapped to the embeddable characters $\mathbf{C}_A$. Due to the specifics of $\texttt{Curve25519}$, when $h_4$ is used, the size of the public keys is $64$ characters. The goal is to transfer (later) LLM-generated messages that embed those $64$ characters in an interactive chat style. For that, in Step 14, Alice needs to generate a random composition $comp_{Alice}$ of the number $64$ into $m$ summands (here, we take $m=10$ summands, but it can be some other suitable values). Each summand is lower bounded by some value (here, we take $low = 3$). According to $comp_{Alice}$ Alice in Steps 17 and 18 computes the partition lists $\mathbf{CPart}_A$ and $\mathbf{bPart_{A}}$. Alice resets her story as an empty string in Step 19. Bob conducts similar computing actions in steps 20 -- 26.

The ephemeral key exchange is initiated by Alice in Step 27 and is followed by the initial response from Bob. Those two initial messages do not contain embedded public key characters. 

Then, in an interactive chat session of $m=10$ messages (back and forward), Alice and Bob gradually embed their public keys, calling the $\textsc{EmbedderLLM}()$ function, trying to continue the stories $Story_{A}$ and $Story_{B}$ correspondingly. The smaller incremental parts of the components of the stories $Story_{A}$ and $Story_{B}$ are depicted as $AliceInput_j$ and $BobInput_j$ for $j \in [0..9]$.
In the final stage (steps 39--42), Alice extracts Bob's chat inputs, extracts his public key, and computes the shared key. Bob performs similar actions in steps 43- 46. The shared keys should be equal, and the value of the $rootkey$ is set to that shared key for some future session.

We can emphasize several properties of Algorithm \ref{Alg:ECDHE-LLM}, which portray the features of our overall framework:
\subsubsection{LLM agnostic} Alice and Bob can use their own locally trained LLMs. The specific local LLMs are invoked in steps 31 and 35. There is no need for those LLMs to be the same.
\subsubsection{Pre- or post-quantum agnostic} While we used pre-quantum elliptic curve public key cryptography as an example, we can easily use some post-quantum key exchange schemes. The expected price will be longer chat sessions for transmitting the ephemeral public keys of the communicating parties.
\subsubsection{Big number of parameter customizations for further mitigation the distinguishing of an encrypted communication} Since the goal is for the chat sessions to conduct undetectable encrypted communication for any totalitarian regime that monitors people's chats, our default parameters $l = 4$, $m = 10$, $low = 3$, as well as the choice of the public key scheme can be additionally chosen by the participants. The choice of the sets of embeddable characters $L_1, L_2, L_3, L_4$ can also be a parameter that can have some variations. Another variant with enriched parameters can be a variant where in between those $m$ Alice-Bob chat-pairs-sentences that hold the embedded characters, there are chat-pairs that are just there for confusion and do not hold any embedded characters.
















\section{Conclusion} \label{sec:conclusion}
In this work, we introduced a novel cryptographic framework that enables secure and covert communication using LLMs. Our approach facilitates encrypted messaging-whether through Public Key or Symmetric Key encryption - while maintaining indistinguishability from natural human-like text in public chat environments. A key advantage of our framework is its LLM agnosticism, allowing participants to use different local models independently. Additionally, it remains resilient against both pre- and post-quantum adversaries, ensuring long-term security. By seamlessly integrating encryption with human-like text generation, our method provides an alternative for secure communication in scenarios where conventional encryption mechanisms are easily detected or restricted. 

Implementing our framework remains as a future work. Furthermore, there are several promising avenues for future research. One immediate direction is to solve the Research Problem \ref{OpenProb:KnownPositions} stated in this work. Another potential direction is the integration of error-correcting codes~\cite{macwilliams1977theory} into our framework. If the bits in the secret positions $\mathbf{b}$ are flipped, the received secret message may deviate from the intended one. Exploring efficient methods to incorporate error correction could enhance the robustness of our approach. Another important direction is optimizing the framework for real-time covert messaging, improving both efficiency and scalability. Investigating broader applications of our framework in privacy-preserving systems could further expand its impact and usability.


\section{Ethical Declaration and Consideration}
The work presented, including theoretical formulations and cryptographic constructions, is original, except for minor language refinements in the introduction and related work sections, where ChatGPT was used. These refinements do not influence the technical contributions or the core results of the paper. This research does not include human subjects, personal data, or ethical risks. 


\bibliographystyle{plain}
\bibliography{reference}  

\onecolumn
\begin{appendices}

\section{Algorithm \ref{Alg:ECDHE-LLM} in a protocol format}
\label{appendix:1}
\begin{figure*}[h!]
    \centering
    \includegraphics[width=1.1\linewidth]{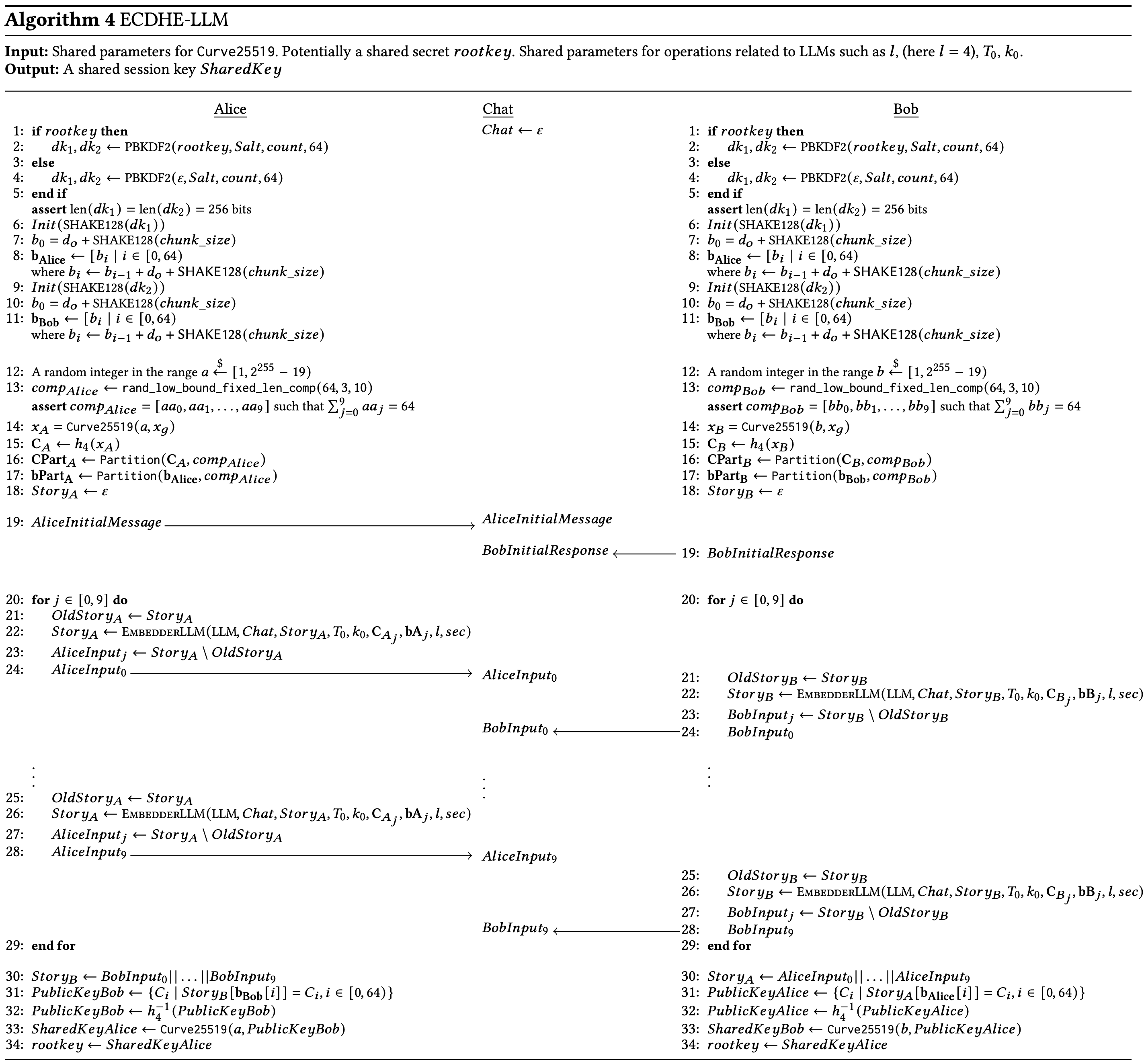}
\end{figure*}


\section{A toy-example of a ECDHE-LLM as a proof of concept}
As a proof of concept for ECDHE-LLM let us use the small didactical elliptical curve \texttt{secp112r1}. Its parameters are the following: $p = ( 2^{128} - 3 ) / 76439$, and the elliptic curve $E$ defined over $\mathbb{F}_p$ is the curve 
\begin{align*}
    y^2 =& x^3 + a_4 x + a_6,\\
    a_4   =& 4451685225093714772084598273548424,\\
    a_6   =& 2061118396808653202902996166388514\\
    G   =& (188281465057972534892223778713752, 3419875491033170827167861896082688).
\end{align*}
Let us assume that Alice and Bob had agreed (on a slightly changed and permuted $L_4$ from the set defined in Section \ref{sec:EmbedderLLM}) set of embeddable characters: $$L_4 = \{\texttt{' '}, \texttt{E}, \texttt{T}, \texttt{A}, \texttt{O}, \texttt{I}, \texttt{N}, \texttt{S}, \texttt{H}, \texttt{R}, \texttt{D}, \texttt{L}, \texttt{C}, \texttt{U}, \texttt{M}, \texttt{W}\}.$$
Using $L_4$, for this toy example, the public keys of Alice and Bob will consist of just 28 characters. Let other shared values be $m = 3$ and $low = 7$. 

\shorten{
Additionally, as a technical convention that is not discussed in Algorithm \ref{Alg:ECDHE-LLM}, let Alice and Bob use the compressed values of the points of the elliptical curves defined with the following two (Python / SageMath) procedures:

\begin{python}
def compress(Point):
    return '0x{0:0{1}X}'.format(2 + int(mod(Point.y(), 2)), 2) + hex(Point.x())[2:]

def decompress(Compressed_Point):
    xx = F(int(Compressed_Point[4:], 16))
    y_square = xx**3 + E.a4() * xx + E.a6()
    yy = y_square.sqrt(extend=False, all=True)[0]
    return E(xx, yy)

\end{python}
}  

Shortening further the text of this example, let us assume that both Alice and Bob have calculated in steps 1--18 the following sequences:
{\scriptsize 
$$\mathbf{b_{Alice}} = [20, 34, 50, 66, 86, 100, 122, 140, 153, 167, 177, 197, 220, 234, 249, 269, 291, 308, 325, 343, 359, 381, 395, 416, 427, 437, 455, 478],$$
$$\mathbf{b_{Bob}} = [15, 40, 51, 64, 88, 99, 113, 134, 152, 176, 198, 222, 245, 267, 291, 302, 312, 332, 355, 368, 392, 417, 440, 454, 468, 483, 508, 531].$$
}

Let Alice randomly generated the following: 
\begin{IEEEeqnarray*}{C}
    a   = 1287024140169629488460400054015049,\\
    comp_{Alice} = [7, 10, 11].
\end{IEEEeqnarray*}
Then, the values of $x_A$, $\mathbf{CPart}_A$ and $\mathbf{bPart_{A}}$ will be 
{\scriptsize 
\begin{IEEEeqnarray*}{C}
    x_A   = \mathtt{0x373ac0abcfdf4cf30b9a0cfbe5d6},\\
    \mathbf{CPart}_A = \mathtt{[ [A, S, A, D, C, \texttt{' '}, D], [L, C, W, U, W, O, C, W, A, \texttt{' '}], [L, R, D, \texttt{' '}, C, W, L, M, I, U, N] ]},\\
    \mathbf{bPart_{A}} = \mathtt{[ [20, 34, 50, 66, 86, 100, 122], [140, 153, 167, 177, 197, 220, 234, 249, 269, 291], [308, 325, 343, 359, 381, 395, 416, 427, 437, 455, 478] ]}.
\end{IEEEeqnarray*}
}

Similarly, let us suppose that Bob has generated the following values:
{\scriptsize 
\begin{IEEEeqnarray*}{C}
    b     = 1447986075431300329725329948154560,\\
    comp_{Bob} = [9, 9, 10],\\
    x_B   = \mathtt{0x7dc525c3afa5307918fb6026d144},\\
    \mathbf{CPart}_B = \mathtt{[ [U, S, D, M, U, S, T, O, W], [M, O, R, L, \texttt{' '}, E, R, A, E], [S, A, U, O, \texttt{' '}, D, H, U, \texttt{' '}, T] ]
},\\
    \mathbf{bPart_{B}} = \mathtt{[ [15, 40, 51, 64, 88, 99, 113, 134, 152], [176, 198, 222, 245, 267, 291, 302, 312, 332], [355, 368, 392, 417, 440, 454, 468, 483, 508, 531] ]}.
\end{IEEEeqnarray*}
}

The following chat conversation resembles what would be a generated chat by Algorithm \ref{Alg:ECDHE-LLM}.  
\begin{figure*}[h!]
    \centering
    \includegraphics[width=1.1\linewidth]{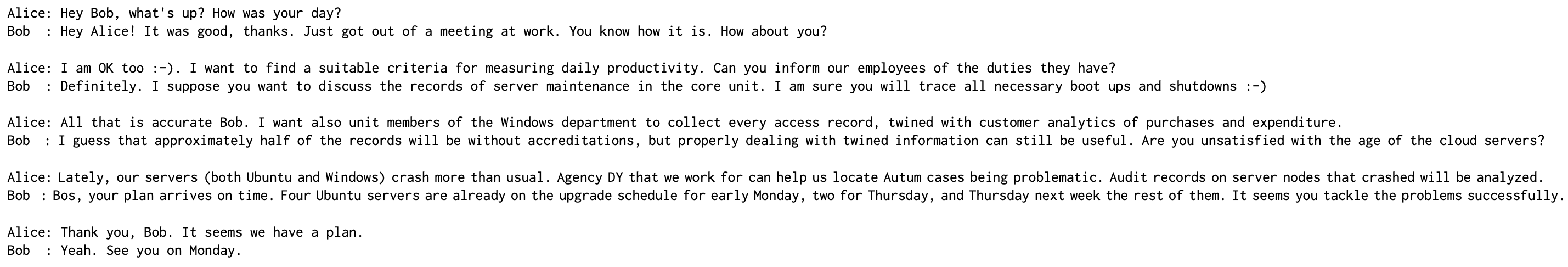}
\end{figure*}

\shorten{
{
\tiny
\begin{verbatim}

Alice: Hey Bob, what's up? How was your day?
Bob  : Hey Alice! It was good, thanks. Just got out of a meeting at work. You know how it is. How about you?

Alice: I am OK too :-). I want to find a suitable criteria for measuring daily productivity. Can you inform our employees of the duties they have? 
Bob  : Definitely. I suppose you want to discuss the records of server maintenance in the core unit. I am sure you will trace all necessary boot ups and shutdowns :-) 

Alice: All that is accurate Bob. I want also unit members of the Windows department to collect every access record, twined with customer analytics of purchases and expenditure. 
Bob  : I guess that approximately half of the records will be without accreditations, but properly dealing with twined information can still be useful. Are you unsatisfied with the age of the cloud servers?     

Alice: Lately, our servers (both Ubuntu and Windows) crash more than usual. Agency DY that we work for can help us locate Autum cases being problematic. Audit records on server nodes that crashed will be analyzed.
Bob  : Bos, your plan arrives on time. Four Ubuntu servers are already on the upgrade schedule for early Monday, two for Thursday, and Thursday next week the rest of them. It seems you tackle the problems successfully.

Alice: Thank you, Bob. It seems we have a plan.
Bob  : Yeah. See you on Monday.
\end{verbatim}
}
} 

\begin{figure}
    \centering
    \includegraphics[width=0.9\linewidth]{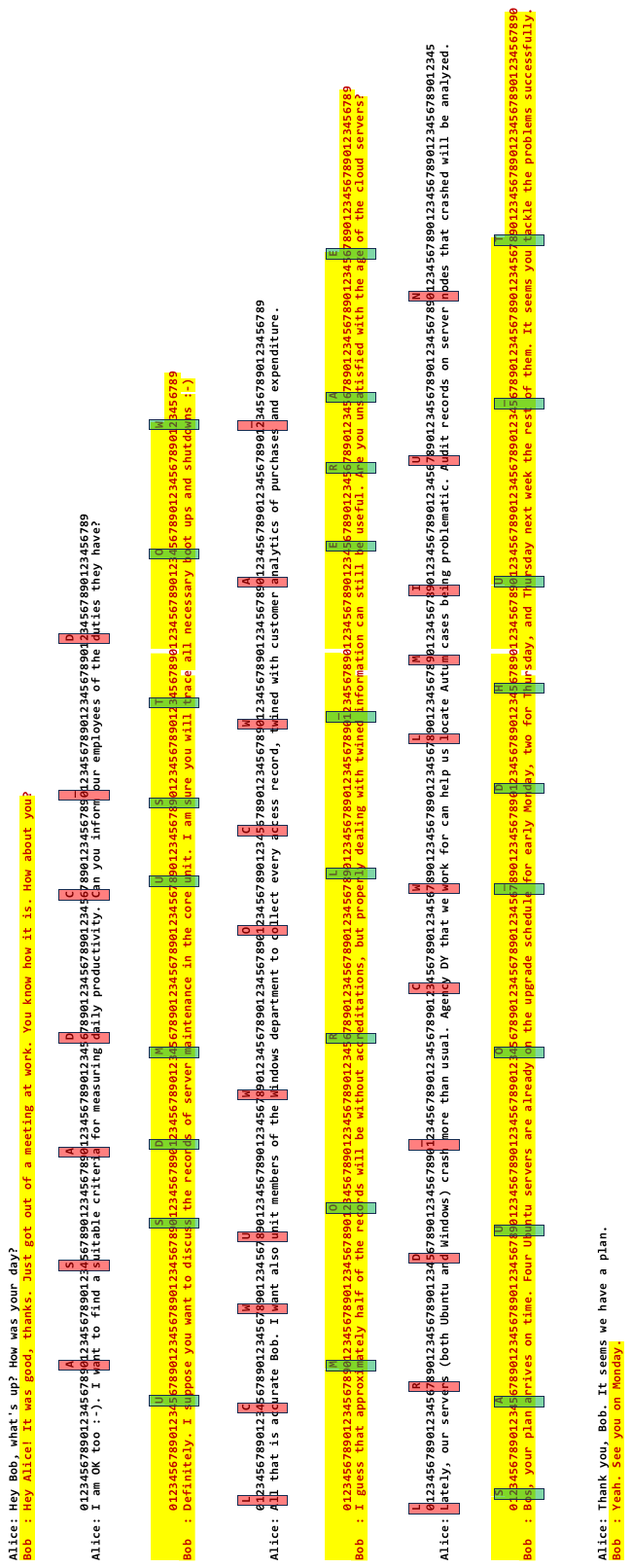}
    \caption{The chat between Alice and Bob with highlighted embedded characters and positions}
    \label{fig:ChatAliceBob}
\end{figure}

\newpage
\section{Adversarial Models and Future Directions} \label{sec:adversarial}

We formalize adversarial models to analyze the security of our framework. These models extend classical cryptographic notions to account for the unique properties of LLM-generated text and the threat of steganalysis.

\subsection{Indistinguishability for Covert Communication (IND-CC)} \label{subsec:ind-cc}

\begin{definition}[IND-CC Game]
Let $\mathcal{A}$ be a probabilistic polynomial-time (PPT) adversary and $\mathcal{C}$ a challenger. The IND-CC game proceeds as follows:
\begin{enumerate}
    \item \textbf{Setup:} 
        \begin{itemize}
            \item $\mathcal{C}$ generates $dk_1, dk_2 \gets \textsc{PBKDF2}(password, Salt, count, 64)$.
            \item $\mathcal{C}$ initializes $\textsc{SHAKE128}(dk_2)$ to sample $\mathbf{b}$ using (\ref{Eq:Sequence_b}).
        \end{itemize}
    \item \textbf{Query Phase:} 
        \begin{itemize}
            \item $\mathcal{A}$ adaptively submits messages $\{m_i\}$. For each $m_i$, $\mathcal{C}$ computes $Enc_i \gets \textsc{AEAD}_{enc}(dk_1, nonce, AD, m_i)$.
            \item $\mathcal{C}$ returns $Story_i \gets \textsc{EmbedderLLM}(\text{LLM}, \textit{TOPIC}, \varepsilon, T_0, k_0, h_4(Enc_i), \mathbf{b}_i, l, sec)$.
        \end{itemize}
    \item \textbf{Challenge:} 
        \begin{itemize}
            \item $\mathcal{A}$ submits $m^*$. $\mathcal{C}$ flips a bit $b \in \{0,1\}$.
            \item If $b=1$, $\mathcal{C}$ returns $Story^* \gets \textsc{EmbedderLLM}(m^*)$.
            \item If $b=0$, $\mathcal{C}$ returns $Story^* \gets \textsc{LLM}(\textit{TOPIC})$ (no embedding).
        \end{itemize}
    \item \textbf{Guess:} $\mathcal{A}$ outputs $b' \in \{0,1\}$.
\end{enumerate}
The scheme achieves \textbf{IND-CC security} if $\mathcal{A}$'s advantage 
\[
\mathsf{Adv}_{\mathcal{A}}^{\text{IND-CC}} = \left|\Pr[b' = b] - \frac{1}{2}\right|
\]
is negligible under Theorem~\ref{Thr:DistinguishinProbabilities}.
\end{definition}

\textbf{Relation to Theorem~\ref{Thr:DistinguishinProbabilities}:} Theorem~\ref{Thr:DistinguishinProbabilities} bounds the probability of tokens deviating from the LLM's optimal parameter ranges ($T \in [0.7, 0.9]$, $k \in [40, 60]$). IND-CC security holds if $\textsc{SHAKE128}$ is pseudorandom and $\textsc{EmbedderLLM}$'s outputs are indistinguishable from natural text (Table~\ref{Tab:Table}).

\subsection{Steganographic Secrecy Against Adaptive Adversaries (SS-ADV)} \label{subsec:ss-adv}

\begin{definition}[SS-ADV Game]
Let $\mathcal{A}$ interact with an oracle $\mathcal{O}$:
\begin{itemize}
    \item $\mathcal{O}.\mathsf{Embed}(m)$: Returns $Story \gets \textsc{EmbedderLLM}(m)$.
    \item $\mathcal{O}.\mathsf{Gen}(TOPIC)$: Returns $Story \gets \textsc{LLM}(\textit{TOPIC})$.
\end{itemize}
$\mathcal{A}$ adaptively queries $\mathcal{O}$ and guesses its mode. The scheme achieves \textbf{SS-ADV security} if $\mathcal{A}$'s distinguishing advantage 
\[
\mathsf{Adv}_{\mathcal{A}}^{\text{SS-ADV}} = \left|\Pr[\mathcal{A}^{\mathcal{O}.\mathsf{Embed}} = 1] - \Pr[\mathcal{A}^{\mathcal{O}.\mathsf{Gen}} = 1]\right|
\]
is negligible.
\end{definition}

\textbf{Alignment with Framework:} SS-ADV security relies on the uniformity of $\mathbf{C}$ characters in $L_4$ (Table~\ref{Tab:FrequencesTop16Characters}) and pseudorandom $\mathbf{b}$ offsets. Adversaries cannot statistically distinguish $h_4(Enc)$-embedded text from natural sequences (see digram frequencies in Table~\ref{Tab:FrequencesTop16CharactersDigramsOffset0}).

\subsection{Public Key Security (ECDHE-LLM)} \label{subsec:pk-dh}

\begin{definition}[PK-DH Security]
Let $\mathcal{A}$ observe a transcript $Chat$ containing public keys $\mathbf{C}_A = h_4(x_A)$ and $\mathbf{C}_B = h_4(x_B)$ embedded via $\textsc{EmbedderLLM}$ (Algorithm~\ref{Alg:ECDHE-LLM}). $\mathcal{A}$'s goal is to compute $\texttt{Curve25519}(a, x_B)$. The scheme is \textbf{PK-DH-secure} if $\mathcal{A}$'s success probability is negligible, assuming:
\begin{itemize}
    \item The hardness of ECDH on Curve25519.
    \item The IND-CC security of $\mathbf{C}_A, \mathbf{C}_B$ embeddings.
\end{itemize}
\end{definition}

\textbf{Relation to Proposition~\ref{Prop:SimpleDistinguisher}:} If $\mathbf{b}$ is leaked (Proposition~\ref{Prop:SimpleDistinguisher}), PK-DH security requires $h_4(x_A)$ to be pseudorandom in $L_4$. Otherwise, security follows directly from ECDH.

\subsection{Token Statistical Analysis (TOK-STAT)} \label{subsec:tok-stat}

\begin{definition}[TOK-STAT Distinguishability]
Let $\mathcal{A}$ know the LLM's token distribution $\mathcal{D}_{\text{LLM}}$ and receive $Story$ generated via $\textsc{EmbedderLLM}$. $\mathcal{A}$ wins if it detects statistical deviations (e.g., via KL divergence). The scheme is \textbf{TOK-STAT-secure} if:
\[
\mathsf{SD}\left(\mathcal{D}_{\text{Embedder}}, \mathcal{D}_{\text{LLM}}\right) \leq 2^{-sec},
\]
where $\mathsf{SD}$ is the statistical distance and $sec$ is the security parameter (Table~\ref{Tab:Table}).
\end{definition}

\textbf{Link to Theorem~\ref{Thr:DistinguishinProbabilities}:} Theorem~\ref{Thr:DistinguishinProbabilities} ensures TOK-STAT security by bounding the probability of sampling tokens outside $T \in [0.7, 0.9]$ and $k \in [40, 60]$.

\subsection{Future Research Directions} \label{subsec:future}

\begin{itemize}
    \item \textbf{Known-Position Resistance:} Address Open Problem~\ref{OpenProb:KnownPositions} by dynamically masking $\mathbf{b}$ or injecting dummy embeddings.
    \item \textbf{Post-Quantum LLM Cryptography:} Integrate lattice-based KEMs (e.g., Kyber) and formalize indistinguishability under quantum queries.
    \item \textbf{Efficiency Optimizations:} Accelerate $\textsc{EmbedderLLM}$'s token search (Steps 19--41, Algorithm~\ref{Alg:EmbedderLLM}) using GPU parallelism.
    \item \textbf{Multimodal Covert Channels:} Extend the framework to image/video LLMs (e.g., DALL--E, Sora) for cross-modal embeddings.
\end{itemize}

\end{appendices}

\end{document}